\documentstyle[preprint,aps]{revtex}  
\begin{document}            
\draft

\input epsf.sty
\input psfig.sty

\title{\bf Quasi-fission reactions as a probe of nuclear viscosity } 

\author{J.Velkovska,
        C.R. Morton$^1$, 
        R.L. McGrath,
        P. Chung,
        I. Di\'{o}szegi}
\address{Department of Physics and Astronomy, State University of New
York at Stony Brook, Stony Brook, NY-11794, USA\\
$^1$ Present address: Department of Nuclear Physics, Research School
        of Physical Sciences and Engineering, Australian National
        University,
 Canberra,Act 0200, Australia} 
\date{\today}
\maketitle
\begin{abstract}
Fission fragment mass and angular distributions were measured from the
$^{64}$Ni+$^{197}$Au  reaction  at 418~MeV  and  383~MeV incident
energy.  A detailed data   analysis was performed, using  the one-body
dissipation  theory implemented in the  code HICOL.  The effect of the
window  and the wall  friction  on  the experimental  observables  was
investigated. Friction stronger  than one-body was also considered.
The mass and angular distributions were consistent with one-body
dissipation.
An  evaporation   code DIFHEAT coupled    to HICOL was
developed in order to predict reaction time scales required to
describe available data on pre-scission neutron multiplicities. 
The multiplicity data were again consistent with one-body dissipation.
The cross-sections for touch,  capture and
quasi-fission were also obtained.

\end{abstract}

\narrowtext
\section{INTRODUCTION} 
\label{sec:intro} 

The mechanism through  which energy is dissipated  in low energy heavy
ion reactions is a  topic widely discussed in the  last 20 years.  The
study  of  deep-inelastic   collisions  in the  1970s   stimulated the
development     of    dissipative    theories  of   nuclear   dynamics
\cite{Hui85,Sch84}.  Following this,  the quasi-fission phenomenon was
discovered in the 1980s. As well as  fusion-fission, it belongs to the
class of the  completely    damped reactions, where all   the  initial
kinetic  energy is dissipated,  but depending on the entrance channel,
the  mass  asymmetry  degree of  freedom  may  be  fully or  partially
equilibrated \cite{Bac85}. Large   mass  transfer  is achieved  in   a
relatively  short  time  (~2-10 x10$^{-21}$~s)   \cite{Tok85} and  the
system quickly re-separates due to the absence of a fission barrier. 

In heavy   symmetric systems, quasi-fission  is the  dominant reaction
mechanism.   A     number of    systematic    studies done     at  GSI
\cite{Tok85,Boc82,She87} using $^{238}$U and $^{208}$Pb beams incident
on  targets  ranging from $^{16}$O  to  $^{89}$Y,  have identified the
following experimental signatures of quasi-fission: 
\begin{itemize} 
\item
Fragment  mass    distributions wider   than the  mass   distributions
resulting from fusion-fission reactions. 
\item
Asymmetries in the mass-angle correlations  increasing with the target
mass. 
\item
Angular anisotropies significantly larger than those in fusion-fission 
reactions.
\end{itemize}

Because of their unique nature, quasi-fission  reactions can provide a
testing  ground   for different models      of dissipative heavy   ion
collisions.  While   in fusion-fission reactions,  the  fragment  mass
distributions  are not  informative    of the reaction  dynamics,   in
quasi-fission, the mass drift towards symmetry is a sensitive probe of
the dissipative  forces  acting  between  the  reaction partners.  The
asymmetry  in    the mass-angle  correlations,   wherever  present, is
directly related to the time the system spends in the coalesced state,
before re-separation. In the  GSI measurements it was determined, that
for  all targets heavier than $^{40}$Ca  the system made less than one
turn before re-separating  and  the turning angle of  the intermediate
complex  was   used  to  extract   the   dynamical  time scale of   the
reaction. The data analysis found that the relevant friction mechanism
is one-body in  nature and is temperature  independent. As a result of
these  works, the extra-push  model  parameters were established in  a
systematic way. They were also used in the development of the one-body
dissipation model implemented in Feldmeier's code HICOL \cite{Fel85}. 
             
Another experimental approach  for determining the  reaction time scale
is   the  measurement of pre-scission    neutron multiplicities.  This
approach is not limited to quasi-fission reactions, but was also used
to  deduce   fusion-fission time scales  \cite{Hin89,Hin92}.  When  the
excitation energy of the coalesced system is  high enough, the neutron
evaporation times are  much shorter than the  scission time and  thus,
the  pre-scission neutron  multiplicity can be   used as a  clock. The
theoretical  treatment of the  data involves a statistical model code,
which is used  to  predict the pre-scission  neutron  lifetimes and to
deduce the reaction time. Although the dynamics of the reaction cannot
be treated with this model, and a direct comparison with the different
dissipation  theories has not been done, the quasi-fission   time scales
derived from pre-scission neutron multiplicities are comparable to the
results  obtained with mass-angle  correlations in similar systems and
temperature regimes \cite{Tok85,Boc82,She87}. 

In a  recent work, Wilczynski  et.al.  \cite{Wil95,Wil96}  re-analyzed
Hinde's et.al.   \cite{Hin89,Hin92}  pre-scission neutron multiplicity
data,  using an  evaporation  code DYNSEQ coupled   to the  outputs of
HICOL.    In that way,   the dynamics of the  reaction  was taken into
account.  The time scales obtained were   an order of magnitude  longer
than the original  results of Hinde  et.al.  Surprisingly large values
for  the  dimensionless   friction coefficient   $\gamma=17-50$   were
reported. These were derived after scaling the one-body dissipation in
HICOL by a factor $k_{s}=4-12$. Using this new analysis, and results of 
analysis   \cite{Hof95} of the GDR   $\gamma$-rays in coincidence with
fission  for   the O+Pb, S+W,Pb   reactions  \cite{But91}, the authors
claimed to observe strong two-body dissipation setting in at nuclear
temperature of about 2~MeV,  with $\gamma$ following $1/T^{2}$ temperature
dependence. One-body friction, for comparison, yields
$\gamma\sim 4$ independent of temperature. 

The strength of two-body dissipation is still an open question in the 
literature. While most authors find that two-body dissipation is
weaker than one-body and yields shorter pre-scission time scales
\cite{Car86,Blo92}, there are some theoretical works that predict strong
two-body viscosity and long time scales associated with this 
(e.i.~\cite{Boi93}). 
Such strong dissipation, if  present, would  influence significantly  the
quasi-fission reaction  dynamics  and the observables  associated with
it:  fragment  mass distributions and   mass-angle correlations. It is
desirable  to  compare dissipation results   deduced from pre-scission
neutron    multiplicities   and  from    fragment    mass  and angular
distributions. At present,  there are four  quasi-fission reactions in
which  the pre-scission   neutron  multiplicities  have been  measured
\cite{Hin92}.  The data set of  mass-angle correlations is larger, but
until now, there  was no measurement  of both observables  in the same
reaction at the  same center of mass (c.m.) energy, and hence, the same
nuclear temperature. 

In this work the mass  and angular distributions  were measured in the
$^{64}$Ni$+^{197}$Au reaction at   the 418~MeV and  383~MeV laboratory
energies.  The mass-angle correlations were also constructed.  For the
first   bombarding   energy, pre-scission   neutron data  exist  and a
comparison of the two experimental approaches is possible.  The second
energy   was used to  investigate   the temperature dependence of  the
friction.  With the development of the code HICOL, a detailed analysis
of  the mass   distributions    became  possible  and    the  relative
contribution of the window  and the wall  friction could be determined
from  the data.  An upper  limit to the reaction
time, and  the friction,   was  determined  by  an analysis    of  the
mass-angle correlations using  the  one-body dissipation model  HICOL.
An   evaporation code DIFHEAT,  coupled  to  HICOL, was developed  and
applied  for  analysis of the pre-scission   neutron data published in
Ref.\cite{Hin92} and a  comparison with Wilczynski's results (obtained
with a different code) is presented.   The deep inelastic, capture and
touch cross-sections were also obtained. 

\section{Experiment}
\label{exp}
The experiment  was performed  at  the Stony  Brook Nuclear  Structure
Laboratory, using $^{64}$Ni beam from the FN Tandem  Van de Graaff and
Super-conducting LINAC accelerators. 

 The reactions, for which the
pre-scission neutron multiplicities were measured and the
two existing interpretations for the  reaction times, determined in Refs 
\cite{Hin92} and \cite{Wil95} are given in Table~\ref{t:table1}.
To complement the pre-scission neutron  multiplicity data, we measured
the mass  and angular  distributions of  the fragments  produced  from
$^{64}$Ni + $^{197}$Au at  418~MeV incident energy. Although, in  this
case the discrepancy shown in Table~\ref{t:table1} is mild compared to the
$^{64}$Ni$+^{238}$U and $^{40}$Ar$+^{238}$U reactions, the  $^{197}$Au
target  was chosen  for  its  lower fissility.   The measurements were
performed  in ``singles'' mode with  Silicon Surface Barrier detectors
(SSBs) and the event reconstruction was done assuming binary fission. 

\begin{figure}
\begin{center}
\vskip 2 cm
\centerline{\psfig{file=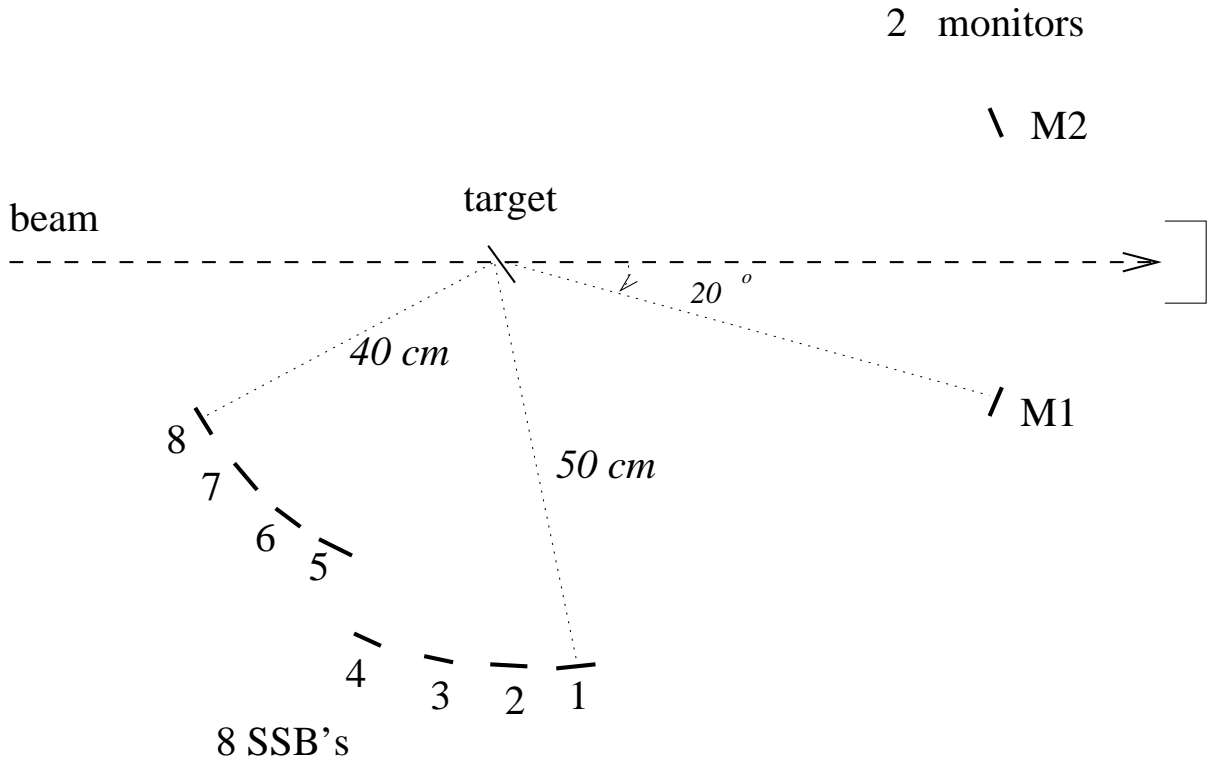,width=16cm}}
\end{center}
\caption{ The experimental layout in the ``BigMac'' scattering chamber.
\label{f:fig1}}
\end{figure}

The experimental layout  in the  2.4~m diameter ``BigMac''  scattering
chamber is  shown in Fig.\ref{f:fig1}.  Eight   large area SSBs (EG\&G
Ortec BF-028-400-60)  were mounted on a  movable platform 10$^{\rm o}$
apart located  at   distances 50   cm and  40   cm from   the  target.
Measurements were done at two positions of the platform, covering from
20$^{\rm o}$ to 90$^{\rm o}$ laboratory angles in the forward position
and from 80$^{\rm o}$ to 150$^{\rm o}$ in the  backward position.  The
platform  was connected  to a  cooling  system and the  detectors were
cooled to $-10^{\rm  o}$C, which allowed over-biasing without damaging
the detectors.  Operating the SSBs in high field strength was crucial,
since  this minimizes  the  pulse-height defect and  the plasma delay.
For the  heavy ions studied in  this experiment, the plasma delay time
was between~$0.2  - 1.5~ns$.  Corrections were  made to compensate for
this  effect.  The pulse height  defects  were  $E_{phd}=5 - 25$  MeV.
Although  systematic  studies of  the two   effects  listed above  are
present   in the  literature  \cite{Boh85,Ogi86},   the  correct  mass
reconstruction for ions as heavy as  the fission fragments measured in
this experiment was   difficult.    An original  procedure for    mass
reconstruction was developed in this work and is discussed in 
Ref.~\cite{Vel98}.  In addition,  two small area ion-implanted silicon
detectors (EG\&G Ortec BU-013-25-300)  were placed at $\pm 20^{\rm o}$
with respect to the beam axis  at a distance  81~cm from the target to
monitor the beam quality and for normalization of the measured fission
fragment yield to the Rutherford scattering cross section. 
 
The  reconstruction of  the  mass and    the energy  of the   detected
fragments,  requires  an  energy and  absolute   Time-Of-Flight  (TOF)
calibration  of the  detectors  and the  subsequent  electronics.  The
energy calibration was  done  using a $^{228}$Th $\alpha$-~source  and
the   elastic  peak   from     a  247~MeV~$^{64}$Ni  beam.    At   the
E$_{lab}=418$~MeV  energy, additional calibration  points from elastic
scattering were  used for the  detectors that were within the reaction
grazing angle.    The   energy  resolution   was $\sim2\%$   for   the
elastically scattered beam and $\sim1.2\%$ for $\alpha$-particles.  An
absolute timing calibration was done  using the reference RF signal of
the LINAC accelerator and the  elastic peak from the  247~MeV run.   A
thin  (100~$\mu$g/cm$^{2}$) self-supporting Au target  was used in the
calibration and  a  250~$\mu$g/cm$^{2}$ Au target  was chosen  for the
measurement, compromising  between  good timing and a  reasonably high
count  rate.   The   width of the   beam  pulse   at   the target  was
$\le~600~ps$ FWHM. Corrections were made  for the plasma delay time of
the heavy fragments, and the overall timing resolution for the elastic
reaction products measured was $\le~1~ns$. 

In addition, the  relative solid angles  of the SSB's and the monitors
were  calibrated  using the   Rutherford scattering   yield.   Several
calibration runs at 247~MeV,    bracketing in time the higher   energy
data, were recorded during the course of the experiment. 
\section{Data Analysis}
\subsection{Mass Reconstruction}
   
\begin{figure}
\begin{center}
\centerline{\psfig{file=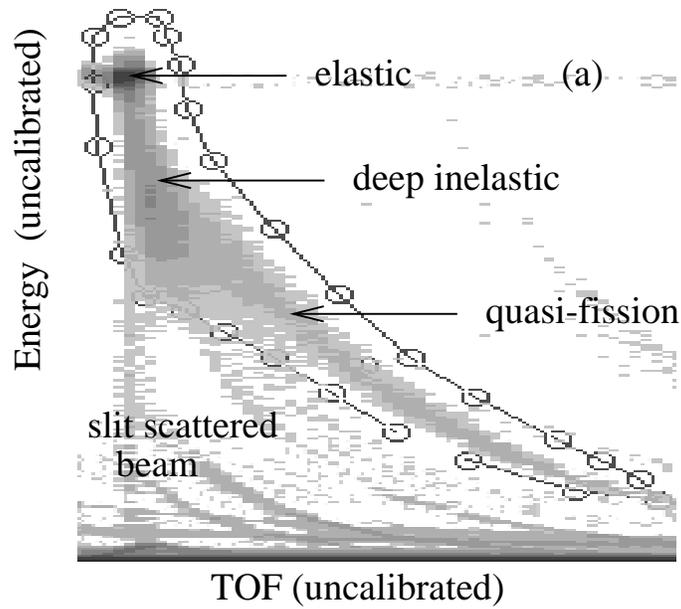,height=8cm,angle=0}}
\end{center}
\begin{center}
\centerline{\psfig{file=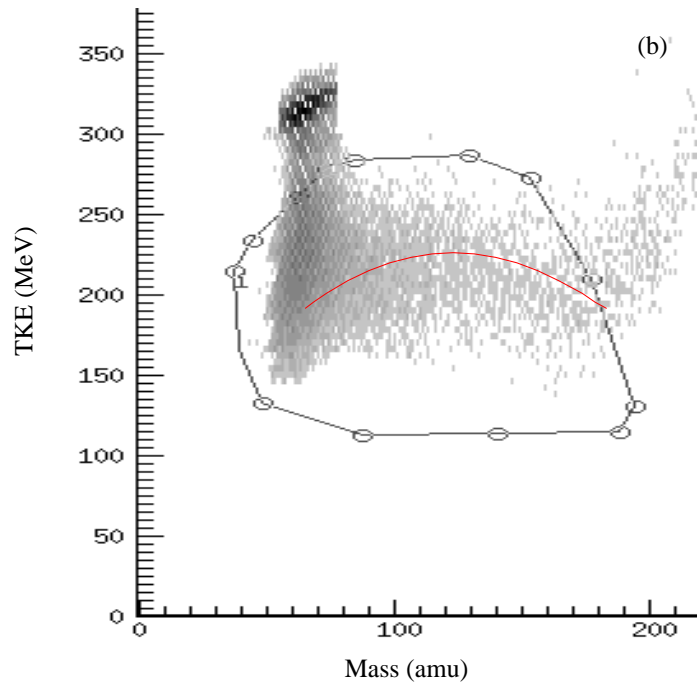,height=9cm,angle=0}}
\end{center}
\vskip -1 cm
\caption{Scatter plots from detector 3 located at 40$^{\rm o}$ in the
lab system for E$_{lab}=418$~MeV. (a) Energy versus TOF 
(uncalibrated). The ``banana'' gate applied to select the 
events of interest is shown. (b) TKE (MeV) versus
Mass (amu). The gate shown separates the quasi-fission and the
deep inelastic events from the elastic and quasi-elastic
scattering. The line shows Viola systematics TKE.
\label{f:fig2}}
\end{figure}

Two-dimensional  (2D) scatter  plots  energy versus TOF were made for
each detector, and the reaction  products were separated from the slit
scattered beam and low-energy-noise  backgrounds by applying 2D gates.
An example of such a plot and the gate applied for detector 3, located
at   $40^{\rm o}$   in the  lab system, is shown in Fig.\ref{f:fig2}(a). 
The high energy elastic peak dominates the yield. Quasi-elastic events
are seen at an energy lower  than the elastics and with  similar TOF. 
The   regions of the deep inelastic  and  the   quasi-fission
fragments are indicated in the figure. 
Although the  elastic and quasi-elastic  channels are not  of interest
for  the quasi-fission reaction studied  here,  these events (wherever
present) provided an important check  for the mass reconstruction,  so
they were included in the first step of the data analysis. Later, when
only  the  damped  reaction  products were  studied,   total kinetic
energy (TKE) versus mass
scatter plots were made  for each detector, and a  gate was applied  to
select the deep  inelastic  and quasi-fission  events only.  See  
Fig.~\ref{f:fig2}(b) for $\theta=40^{\rm o}$ (lab). 
 
After selecting the events of interest, the  mass was reconstructed, as
described in Ref~\cite{Vel98}, with overall mass  resolution better than 10
mass units. 

\subsection{Efficiency Simulations}
\label{s:simu}
\begin{figure}
\begin{center}
\centerline{\psfig{file=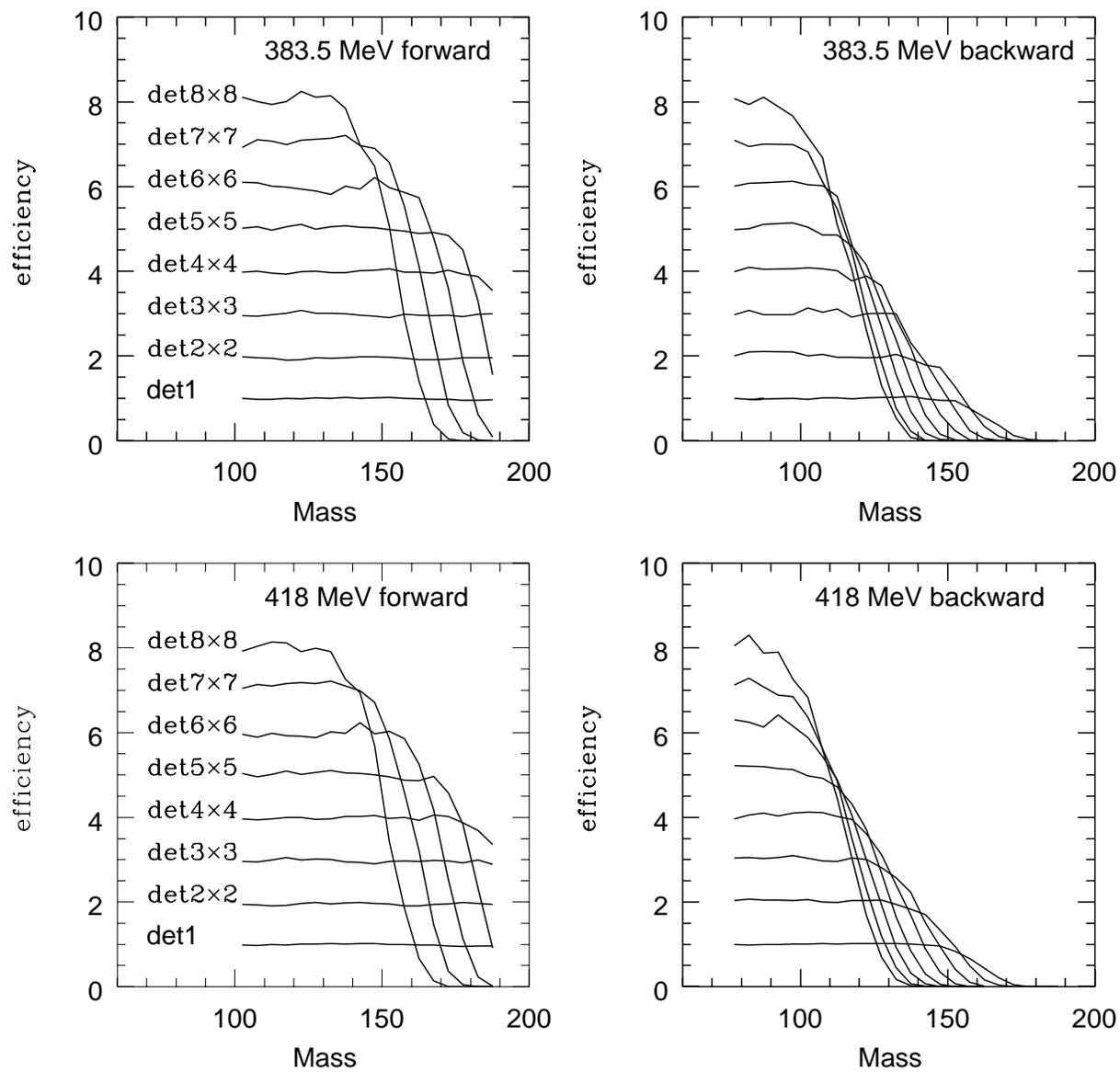,width=16 cm }}
\end{center}
\caption{Efficiency curves for all 8 detectors in the two positions of the
platform for E$_{lab}=418$ MeV and E$_{lab}=383$ MeV.\label{f:fig3}}
\end{figure}

A Monte-Carlo  simulation  program  was  developed to   determine  the
detection efficiency for each mass at each  detector position.  Events
were   generated in  the    c.m.   with  flat  mass  distribution    and
$d\sigma/d\Omega_{c.m.}=1/\sin\theta$ (assuming  isotropic  distribution
in the polar angle $\theta$). The  kinetic energy of the fragments
was drawn from a Gaussian distribution with parameters determined from
the measured TKE distributions, as discussed later. 
The simulations  included energy loss in  the target  and in the front
$Au$ electrode of the detector,  pulse-height-defect and plasma delay,
electronics energy thresholds,  TOF and velocity cuts.  After applying
the   experimental cuts and    the  experimental mass   reconstruction
procedure, the  events  were  transformed back  into   the c.m.  and the
``detected'' yield was compared to the input. The efficiency curves for
the two  energies studied and  for the two  positions  of the platform
holding  the  detectors  are shown in  Figure~\ref{f:fig3}.  For small
masses, the input angular  distribution is recovered from the ``data'',
in the   intermediate   mass  region   (A$=125-130$),  the   detection
efficiency  is reduced in the three  most backward detector positions,
and the heavy fragments (A$>150$) can only  be detected in the forward
position of the platform  (from $20^{\rm o}$  to  $90^{\rm o}$ in  the
laboratory system). 

The  mass cuts observed  in this  experiment  agree very well  with the
simulated   efficiency curves.  This  gives confidence  that  the mass
reconstruction in the experimental data was  done correctly, and also,
that all relevant effects were  included in the simulations. Based  on
the simulated efficiency, correction to   the experimental yield   was
done for the mass bins in which the  efficiency was between 0.5 and 1.
For lower efficiencies, the  centroid  of the fragment  kinetic energy
distribution  is not  covered, so  these  data were excluded from  the
analysis. 

\section{Results and Discussion}   
\subsection{Fragment Mass-TKE Distributions} 

2D scatter plots  of TKE (in the  c.m.)  versus Mass  were generated for
each  detector.  The   data   were corrected    for  pre-scission  and
post-scission neutron   emission, using  the measured  quantities from
Ref.~\cite{Hin92}:  for   418 MeV,  $\nu_{pre}=3.15$, $\nu_{post}=2-5$
depending   on the   mass  of the   fragment.   For E$_{lab}=383$~MeV,
experimental data are not available, so values  estimated on the basis
of  HICOL's predictions for the heat  along the trajectory and in each
fragment  at   the  scission  point were    used:  $\nu_{pre}=1.5$ and
$\nu_{post}=1-3$ increasing with the fragment mass. In 
Fig.~\ref{f:fig2}(b)  an example of TKE-Mass   scatter plot is shown.
The  line  indicates   the Viola-systematics   energy.   The  measured
centroids of the  TKE distributions agree  with  Viola systematics and
with   previous  quasi-fission  studies \cite{Tok85,Boc82,She87}.  The
variance of the distribution,  for which systematic data and empirical
formulae do not  exist, was determined to  be $\approx 15\%$ from  the
centroid   and  is  again   consistent  with    the measurements    in
Ref. ~\cite{She87}.  The HICOL calculations,  discussed later, predict
correctly the centroid of  the TKE distribution,  which shows that the
friction mechanism implemented in  the code produces the right amount
of damping. 

\subsection{Fragment Angular Distributions} 
\label{s:angdist} 
\subsubsection{Experimental Results}
The energy-integrated cross-sections binned in 10 mass units,
$\frac{d\sigma}{d\theta_{c.m.}}= 2\pi \sin
\theta_{c.m.}\frac{d\sigma}{d\Omega_{c.m.}}$, were constructed, after
lab to c.m. transformation and normalization to the Rutherford scattering
yield. 

Figures~\ref{f:fig4} and \ref{f:fig5}  show the  results from the  two
energies measured. The  data for the masses up  to  the symmetric mass
split are  shown with  full  points.  Above symmetry,  only a  limited
angular range was covered, due  to the various factors restricting the
measurement  and the  mass reconstruction. The   data obtained in this
region were used for  a  consistency check and  are plotted  with open
points in their complementary mass bins.  The  error bars shown in the
figures represent  the  statistical errors from  the  measurement. The
data  show  a  significant forward  - backward   asymmetry typical for
short-time scale reactions in which the rotational period of the system
is longer  than the reaction  time.  This behavior is  consistent with
one-body friction.

\subsubsection{HICOL Calculations Along the Trajectory}
\label{s:trj}
A  series of  HICOL calculations   was  performed to investigate   the
dissipation mechanism responsible  for the forward-backward  asymmetry
in  the experimental angular   distributions.  The  code  implements a
macroscopic  description of     dissipative  heavy   ion    collisions
~\cite{Fel85}. A set of   collective shape and angular coordinates  is
used  and  their evolution   is  followed, solving  the   equations of
motion. The coupling of  the macroscopic and  the intrinsic degrees of
freedom is considered explicitly in the derivation of the friction and
the mass diffusion  tensors, in which Fermi  gas one-body formulae are
used.  A  typical ``along-the-trajectory''  calculation, which will be
used  as an illustrative  example is  shown in  Fig.  \ref{f:fig6}. We
will consider a quasi-fission  reaction of $^{64}$Ni$+^{197}$Au at 418
MeV    laboratory energy    and   initial orbital   angular   momentum
L$=50\hbar$.  The potential energy, the  kinetic  energy in the  shape
degrees of freedom  T$_{s}$,   in  the  angular degrees    of  freedom
(T$_{rel}$  and T$_{int}$) and the  dissipated energy are plotted as a
function of time. In panel (a), the actual shapes along the trajectory
are  also shown.  The trajectory starts  at time $t=0$ when the nuclei
are 14~fm apart.  All the  energy  available to the system  resides in
kinetic energy   of     the shape   and  relative    rotation.  Within
6x~$10^{-22}s$, the nuclei   come  into contact.  At  that  time,  the
proximity friction has already produced about 8 MeV of heat. After the
touching  point,  as the window opens,   the radial motion  is rapidly
damped.  With this, the shape loses most of its kinetic energy, since
the  other two shape  coordinates - the neck  and the asymmetry do not
possess   any initial momentum.  The viscous   drag in the window sets
into motion the intrinsic rotation. 
At $t=15$x$10^{-22}s$ the  window has produced 68 MeV  of  heat and is
already widely open. At this point  the shape changes from a necked-in
to a convex  type. The system  enters  the mono-nuclear regime, and  a
uniform  wall is considered   for  the subsequent evolution.    A long
creeping  motion  with  practically no kinetic    energy in the  shape
follows. Without  the window friction against   the net particle flux,
the asymmetry degree of freedom  begins relaxing. It has been severely
hindered up to now. As  we move along  the trajectory, the deformation
develops again, the window  comes    back and the  total   dissipation
increases. The potential  is  rapidly falling,  the system gains  back
kinetic energy in  the shape (10  MeV in this  case)  and it scissions
separating with  final  angular  momentum L$_{f} =43\hbar$  and  final
fragment mass numbers A$_{1}=137$  and   A$_{2}=124$. The TKE of   the
fragments is  221 MeV, consistent with  our measurement and with Viola
systematics. The mass variance
is $\sigma_{AA}^{2}=888$ and  the total number of exchanged  particles
through    the window~- N$_{ex}=6614$.  Within  the   reaction time of
~$\simeq 140\times10^{-22}s$ the  system  does not complete   one full
rotation ( the scattering angle is $-54^{\rm o}$). The fragment masses
are  correlated with   the  emission  angles,  which will   result  in
asymmetric mass-angle  distributions.  This is a typical quasi-fission
trajectory, where the system  scissions bypassing the compound nucleus
stage. The mass asymmetry  is not fully  relaxed,  but almost all  the
initial  kinetic energy is  dissipated   and the fragments fly   apart
driven by the strong Coulomb repulsion. 

In the model implemented in HICOL, there  are no free parameters.  The
only    {\it{ad  hoc}} procedure   used   is  in   the combination  of
window-plus-wall and  mono-wall friction. For  shapes with a ``fat'' neck,
there is an ambiguity in  the choice of a  window plus two walls, or a
mono-nucleus  shape  without a  window.   To make  a smooth transition
between  the two regions, a  form factor $f(x_{n})$   multiplying the rate of
energy dissipation $\dot{Q}$, is introduced as follows: 
\begin{equation}
\dot{Q}=(1-f(x_{n}))\dot{Q}^{win+wall}+f(x_{n})\dot{Q}^{mono},
\end{equation}
with $x_{n}=R_{neck}/min(R_{1},R_{2})$ measuring the ratio between the
neck radius and the radius of the  smaller nucleus. For ``thin'' necks
($0<x_{n}<0.8$),  $f(x_{n})=0$  and the full window-plus-wall friction
is used.  In the region $0.8\leq x_{n} \leq  1$, where the shape makes
the  di-nucleus to  mono-nucleus  transition, $f(x_{n})$   is smoothly
rising from  0   to  1 and   both widow-plus-wall   and  mono-wall are
used.  For $x_{n}>1$,  the energy dissipation   is calculated with the
mono-wall formula. In the outgoing part of the trajectory, when the
neck develops again, $f(x_{n})$ drops from 1 to 0, after which, the full
window-plus-wall friction is used until the scission point. The
functional form of $f(x_{n})$ and the limiting value $x_{n}=0.8$ are 
arbitrarily chosen. The  concern when using  such a  description comes
from the fact  that the two formulae  ( window-plus-wall or mono-wall)
give  different     results, when  applied  to  the    same  shape. The
window-plus-wall formula  can produce 4  to 8 times  stronger friction
against the mass-asymmetry equilibration than  the wall formula alone.
Fig.~\ref{f:fig7}~(a)  illustrates the mass evolution of the
projectile-like and the target-like fragment along the
E$_{lab}=418$~MeV, L$=50~\hbar$ trajectory. The evolution of the form
factor $f(x_{n})$ is shown below it in Fig.\ref{f:fig7}~(b). One can
easily see, that the main mass transfer from the target to the
projectile happens during the mono-nucleus part of the trajectory,
when the window friction is turned off. In Fig.~\ref{f:fig7}~(c)  and
(d) we have calculated the mass evolution in the two extreme cases:
mono-wall friction along the whole trajectory, or full
window-plus-wall everywhere, respectively. Both these cases are clearly
unphysical. Without the window friction, all trajectories proceed
rather rapidly to symmetry, which is in contradiction with the
experimental results (Section~\ref{s:mdist}) , while with the
window-plus-wall acting everywhere, there are no trajectories that produce
symmetric mass-splits. The realistic description of the reaction
dynamics needs a mixture window-plus-wall and mono-wall friction. We 
consider that the form factor $f(x_{n})$ should be used as free
parameter and adjusted to match the experimental mass distributions.

In the dynamical analysis of  Hinde's et.al. pre-scission neutron data
\cite{Hin92},      Wilczynski, Siwek-Wliczynska   and         Wilschut
\cite{Wil96} scaled the  energy dissipation rate $\dot{Q}$ by an
additional factor $k_{s}$, which was treated as a free parameter.  The
factor $k_{s}$ was allowed to obtain values smaller than unity, when a
reduction of the full one-body  friction was needed, and $k_{s}>1$ was
interpreted  as    evidence for    two-body  dissipation.   Along  the
trajectory,  the  authors of   Ref.\cite{Wil96} divided   $k_{s}$ into
$k_{s}^{in}$ and $k_{s}^{out}$. In  the  early stage of the   reaction,
when the system is   reasonably cold,  a factor $k_{s}^{in}=0.5$   was
used, and after  the mono-nucleus stage  is reached and the system has
converted    a significant amount   of    kinetic energy into  heat  -
$k_{s}^{out}=4 - 12$ was applied to make the reaction time long enough
to  allow   the  emission of  the   requisite  number of  pre-scission
neutrons. The justification for using such a description was the
expected temperature dependence of nuclear viscosity.  

Fig.\ref{f:fig8} shows  the  L=$50\hbar$  trajectory  calculation  for
$^{64}$Ni$+^{197}$Au  at E$_{lab}=418$~MeV,   which   was done   using
$k_{s}^{in}=0.5$ and $k_{s}^{out}=4$ as described in Ref.~\cite{Wil96}. 
The
quantities plotted are  the same  as  in Fig.\ref{f:fig6}, which was
for the 
standard   ($k_{s}=1$)  HICOL   trajectory.   By  comparing  the   two
calculations, one sees that the shape  of the potential energy, kinetic
energy   and  total heat curves   is   similar, but  the  time scale is
stretched in the large friction trajectory. The relative contributions
from window and  wall dissipation are different.  By  scaling down the
friction in the early stage  of the reaction when the window-plus-wall
formula is used, the  window  contribution is  reduced. Then, in   the
mono-nucleus stage, the weaker wall friction is scaled  by a factor of
4. This brings   the  system into  a relatively  flat  region  of  the
potential  with very  little  kinetic energy (mainly  in rotation) and
strong friction hindering its  motion  towards scission. It takes  3.5
times longer to develop the deformation and find its  way to the steep
slope   of the potential  that  leads  to scission.  During this  long
reaction time, the system makes more than  two full rotations, as seen
in the shape evolution in the  figure. The calculated final scattering
angle    is     $-760^{\rm o}$.    This will    lead     to  orbiting type
($\sim1/\sin\theta$) angular distribution of the fragments, instead of
the  asymmetric   distribution  expected   from  the   trajectory  in
Fig.\ref{f:fig6}, which agrees qualitatively with the experimental results. 
The mass drift   in the large-friction trajectory  is   similar to the
one-body    case, since  the window  friction   was initially  reduced
($k_{s}^{in}=  0.5$), and  later increased,  in the final  part of the
trajectory. 

\subsection{Cross-Sections} 

\begin{figure} 
\begin{center}
\centerline{\psfig{file=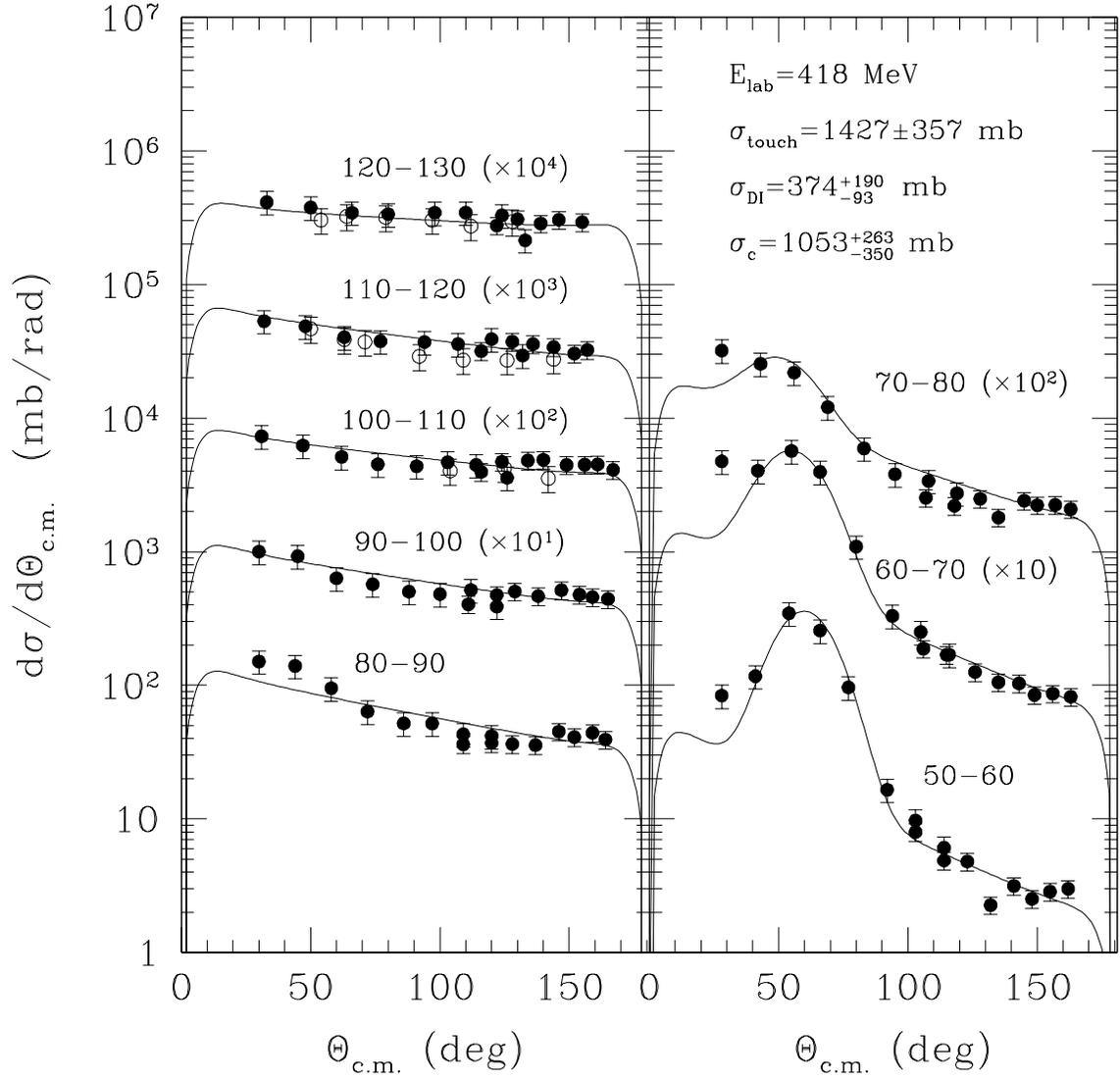,height=16cm,angle=0}} 
\end{center}
\caption{Fragment angular distributions in bins of 10 mass units from the 
E$_{lab}=$418~MeV measurement. The fits to the data are described in the text.
The values of the measured cross sections are given. The open points come
from the measurements in the complementary mass bin with the assumption that
binary fission is observed.\label{f:fig4}}
\end{figure}

\begin{figure} 
\begin{center}
\centerline{\psfig{file=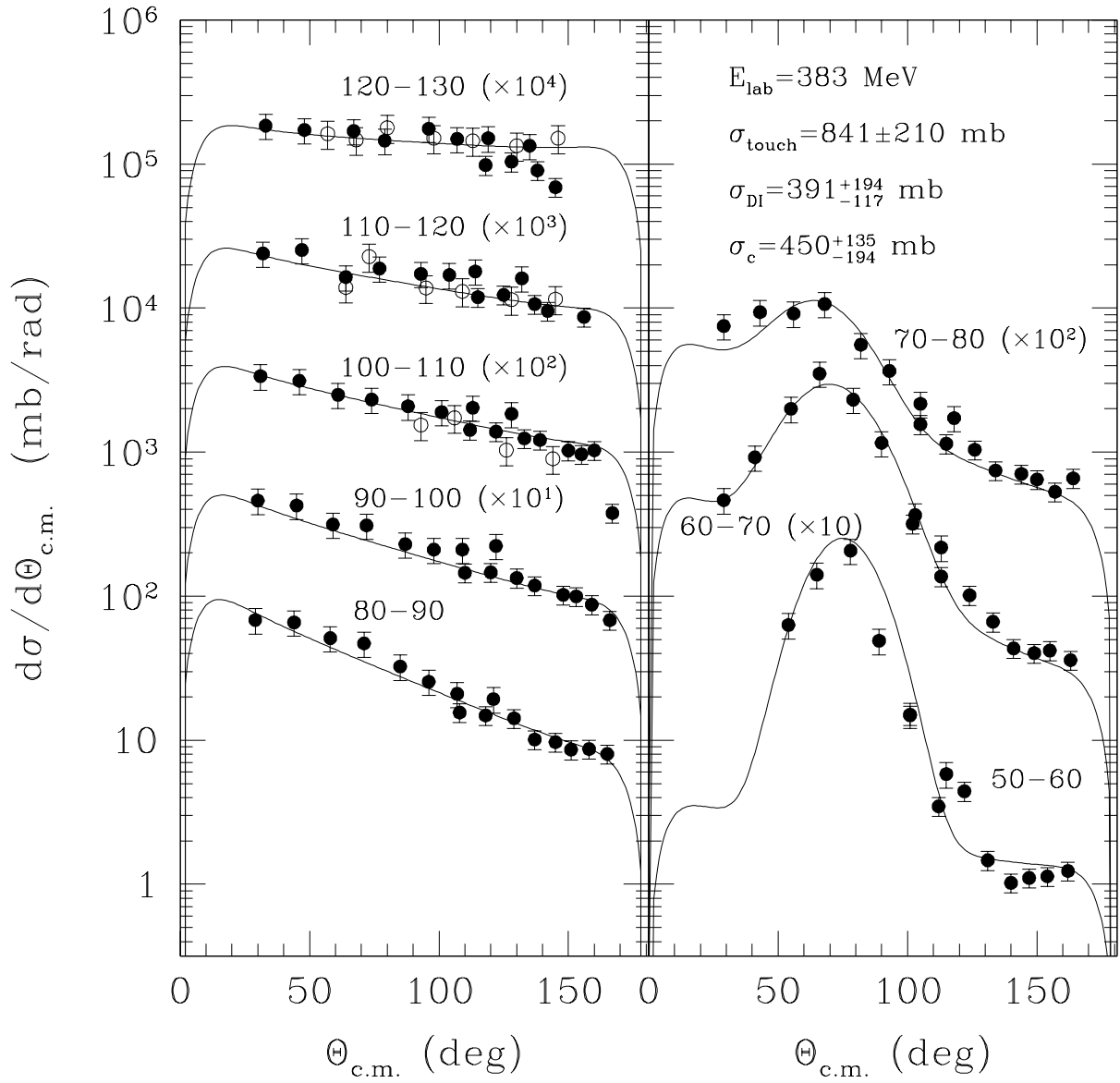,height=17cm,angle=0}} 
\end{center}
\caption{As per Fig.\ref{f:fig4} but for E$_{lab}=383$ MeV.
\label{f:fig5}}
\end{figure}

The experimental angular distributions were extrapolated to $0^{\rm o}$
and $180^{\rm o}$ 
as described by Back et.al. in Ref.~\cite{Bac96}.
For fusion-fission reactions the angular distributions of the fragments are
given by \cite{Van73}:
\FL
\begin{eqnarray}
\label{eqn:fiss}
W(\theta)=&&\sum_{I=0}^{\infty}(2I+1)T_{I}\nonumber\\ 
&&\times{\frac{\sum_{K=-I}^{I}{\frac {1}{2}}
(2I+1){\left|d_{0K}^{I}(\theta)\right|}^{2}\exp\left[-K^{2}/2K_{0}^{2}\right]}
{\sum_{K=-I}^{I}\exp\left[-K^{2}/2K_{0}^{2}\right]}}
\end{eqnarray}                                
Here $I$ is   the total spin of  the  fissioning nucleus,  $K$  is the
projection  of $I$  on the  fission  axis, $T_{I}$ is the transmission
coefficient   for   fusion   of     the  $I$th  partial    wave    and
$d_{0K}^{I}(\theta)$   are    the   real    parts   of   the    Wigner
$D$-functions. Eq.     \ref{eqn:fiss} gives    angular   distributions
symmetric    about $90^{\rm o}$   which   approach the  $~1/\sin\theta$
classical distribution for   $I>>0$.    These are not    applicable to
describe  the  asymmetric   angular  distribution  in  short-time scale
quasi-fission reactions. However, since a theoretical prescription for
this type reactions is  not  available, Back et.al. \cite{Bac96}  used
the following expression to fit the data from a measurement similar to
the one done in this work: 
\begin{equation}
\label{eqn:angd}
\frac{d\sigma}{d\theta_{c.m.}}=2\pi \sin \theta_{c.m.}
\exp(\beta(\theta_{c.m.}-\pi/2))W(\theta_{c.m.}),
\end{equation}
where  $\beta$ is the angular slope  parameter.  The same approach was
adopted  here.  The transmission  coefficients $T_{I}$ were calculated
using the extra-push  estimate  to the complete   fusion cross-section
with a program due to Back \cite{Back}.   The Wigner $D$-functions for
$I,K\leq 25$ were calculated  using the  CERN library routine  DDJMNB,
and the   recursion  relations   \cite{Var88} were  used   for  larger
spins.  The value of $K_{0}$ was  determined from the mass bin closest
to symmetry, and kept fixed in the other mass bins.  Best results were
obtained  with  $K_0^2=25^{+25}_{-10}$   for  the  418~MeV    data and
$K_0^2=15^{+20}_{-5}$ at 383~MeV.   The overall normalization  and the
parameter $\beta$ were varied to produce best fits to the data. In the
systems studied by Back et.al.\cite{Bac96} the clean separation of the
quasi-fission  and deep inelastic events was  possible, since the mass
distributions were peaked at symmetry.  In  the case studied here, the
mass distribution  has a  minimum at  symmetry  and these two reaction
channels could not  be separated (see  also Fig.\ref{f:fig2}. The deep
inelastic  scattering exhibits a maximum  at the  grazing angle of the
reaction  for $A  < 80$ (right   panels  in Figures ~\ref{f:fig4}  and
\ref{f:fig5}) and  the above description  of  the angular distribution
does not  apply. The laboratory  grazing angles for E$_{lab}$  418 MeV
and 383 MeV are $55^{\rm o}$ and $65^{\rm o}$, respectively.  In order
to fit the data, a Gaussian distribution was added to the distribution
given  in Eq.\ref{eqn:angd} and the relative  contributions of the two
distributions    were  adjusted   to    obtain   agreement   with  the
measurement.  With this,  an  attempt was  made  to estimate the  deep
inelastic and  the capture cross-sections  separately.  For the $A\geq
80$ mass bins (left panels in Figures ~\ref{f:fig4} and \ref{f:fig5}),
the cross section is only due to capture reactions. The values for the
total                cross           section           measured,
$\sigma_{touch}=\sigma_{DI}+\sigma_{capture}$,     and  the  estimated
contributions   from    the   two reaction  channels     are  given in
Figs.4,5. The errors in $\sigma_{touch}$, around $25\%$, come from the
quality  of the fits and  the  uncertainty in determining the  monitor
position and the detector  solid angle  ratios. The contribution  from
the statistical errors  of the data is  much smaller.  The upper limit
of the  deep inelastic cross section  $\sigma_{DI}$ was estimated with
the assumption that for $A < 80$  all the cross-section forward of the
grazing angle   is in  this  channel.  This  brings  the corresponding
reduction   in   the    capture cross-section.   The  angle-integrated
cross-sections for the   different   mass bins are  listed   in Table~
\ref{t:table2}. 

\subsection{Mass Distribution}
\label{s:mdist}

After integrating   the  angular distributions in the   different mass
bins,  the   mass distributions for   the  two energies  measured were
obtained (see Fig.~\ref{f:fig9}).  At both energies, the distributions
differ  dramatically from  the   mass distributions of  fusion-fission
reactions, which  are narrower and peaked   at symmetry. This behavior
can   be  qualitatively explained since,    for this system  which has
fissility  $x=0.915  $, the   symmetric  saddle  does not  exist.  The
potential  energy landscape  always  favors asymmetric  mass division.
Theoretical mass distributions were obtained using HICOL's predictions
for  the final  mean  fragment  masses and   their variances for  each
trajectory that leads  to  a damped   reaction.  A weighted sum   over
angular momentum was constructed,  using the measured  cross-sections.
The  resulting  mass  distributions are   plotted   together with  the
experimental data in  Fig.~\ref{f:fig9}.  The  dashed-line histograms
give  the HICOL  predicted  mass-distributions.  Although the  general
trend  of the  data is  reproduced,  a  quantitative agreement is  not
achieved. The  code under-predicts  the deep inelastic  scattering and
over-predicts  the  yield at the   symmetric  mass division  for  both
energies. Since the mass drift is governed by the window friction, and
in HICOL the  relative contributions of window-plus-wall and mono-wall
friction are mixed with a form factor determined from the shape of the
system along  the trajectory, as discussed above,  a variation  of the
form  factor  was considered in this  work,  in order to reproduce the
experimental  results. Allowing for the  window friction to contribute
for  larger $R_{neck}$   by   changing  the  limits  of  $x_{n}$  from
$0.8\leq x_{n}\leq 1$ to $0.95\leq x_{n}\leq 1$ gave a satisfactory result for
the 418~MeV data and  a trend in the  right direction for the  383~MeV
data  (the solid-line   histograms in the   figure). Further 
increase in the lower limit of $x_{n}$ still did not reproduce the 
large deep inelastic cross section measured at E$_{lab}=383$~MeV.
                  
The  discrepancy between  HICOL  and  the  data in  the deep inelastic
channel was addressed by Feldmeier \cite{Fel85}  and was attributed to
the fact that  the  narrowly-peaked phase  space approximation  is not
strictly valid for the deep  inelastic trajectories. The mass variance
is under-predicted in this region, which limits the partial waves that
can contribute to the cross-section in  the near-projectile and target
mass    bins.  Including fluctuations  in  the   trajectory model will
probably  improve  the  agreement  with the data.   The  code  gives a
reasonable  description  of the  418~MeV   data,  in which   the  deep
inelastic   events   comprise  $\approx  1/4$     of  the total  touch
cross-section. At the  lower energy, where  almost  half of  the touch
cross-section is in the deep inelastic channel,  only the shape of the
distribution is reproduced by the calculations. 

Comparison of the E$_{lab}=418$~MeV mass  distribution data with other
theoretical   descriptions from     the literature  is   presented  in
Fig.~\ref{f:fig10}.    In the work  by Shen  et.al.\cite{She87}, it was
found  that the full  window-plus-wall dissipation gave reaction times
which were too long and inconsistent  with the time scales derived from
the  mass-angle correlations. The authors came  to the conclusion that
the window friction should not be included in the calculation. 

In Fig.~\ref{f:fig10}~(a), a calculation without the window friction is
presented.  It differs  dramatically from  the data,  since the window
friction is the main mechanism that prevents the fast equilibration of
the  mass asymmetry.  Fig.~\ref{f:fig10}~(b),  shows a calculation done
using Wilczynski's \cite{Wil96} prescription  for scaling the friction
along  the  trajectory,   which  was   already   discussed in  Section
~\ref{s:trj}.  Although   this   calculation   reproduces   the   mass
distribution, it does not reproduce the mass-angle correlations, since
the reaction times are longer than the rotational period of the system
(See Section~\ref{s:trj}). 

A  frequently  used     approach to  implement  friction   in  various
descriptions of heavy-ion   reactions   is  to   use  window-plus-wall
dissipation   during      the  whole  reaction,    but    scaled  by a
factor$\approx0.3$, which   was determined  from fitting  experimental
fission    fragment    TKE   distributions     and    giant  resonance
widths~\cite{Nix87}. Figure~\ref{f:fig10}~(c) shows the calculated mass
distribution using this  approach, in comparison  to the 418~MeV data.
With the window friction present during the  whole trajectory, even if
it  is  scaled  down, the mass   drift  towards  symmetry is  hindered
significantly, resulting in disagreement with the data. 

Using a shape dependent friction along the whole trajectory, as it was
done   in  HICOL,  reproduces  in   a self-consistent    way  the mass
distributions (Fig.\ref{f:fig9}(a)),     the   mass-angle correlations
(Figs. ~\ref{f:fig4},\ref{f:fig11}) and  the fission fragment TKE. The other
approaches discussed  here,    fail to reproduce all   observables  at
once.  There   are  minor discrepancies   between the   model  and the
experimental  results  which  show   that  the   form   factor mixing
window-plus-wall and mono-wall friction requires further investigation.
The mass   variances for the   deep inelastic trajectories   also need
further  refinements in order   to   yield better agreement  with  the
experimentally measured cross-sections. 

\subsection{Mass-Angle Correlations and Reaction Times}
\label{s:ma}
From the measured angular distributions extrapolated to $0^{\rm o}$ and
$180^{\rm o}$, the double differential
cross-sections $d^{2}\sigma/(d\theta_{c.m.}dA)$ were constructed.
 
The results are plotted in  Fig.~\ref{f:fig11}. The contour levels are
listed in the figure.  A forward-backward asymmetry, which reveals the
short time scale of the reaction, is  clearly evident for both energies
measured.   A two-body    friction  mechanism with $1/T^{2}$
temperature dependence, as suggested in
Ref.~\cite{Wil96}, would require friction scaling factors of
$k^{out}_{s}=10-12$ for E$_{lab}=383$~MeV  and $k^{out}_{s}=4$ for the
higher energy, which in both cases results in time scale much longer
than the rotational period of the system and flat
$d\sigma/d\theta_{c.m.}$ distributions. Only  the   deep   inelastic
component   of the  {\it    {touch}}  cross-section would  retain  the
asymmetry about $\theta_{c.m.}=90^{\it o}$.  Since the data do not  show
this behavior,  the large friction  scenario is {\it  {ruled out}}.  A
direct  comparison of the  measured double differential cross-sections
$d^{2}\sigma/(d\theta_{c.m.}dA)$ and HICOL could not  be done, since the
code predicts the  mean  value of the scattering  angle,  but does not
calculate its variance. 

For   all partial  waves, the   calculated  mean scattering angles are
smaller than $180^{\rm o}$, which  is consistent with the experimental
observation of  a  non-orbiting angular distribution.   Although HICOL
suggests  the  correct behavior,   there is  an  indication  that  the
reaction times for the intermediate  masses are over-predicted,  since
in the experiment, all fragments in this range are peaked forward with
a maximum yield angle in  the vicinity of  $20^{\rm o}$ and an average
angle of $60^{\rm o}-80^{\rm o}$. A  larger saturation angle ($\approx
100^{\rm  o}$)  is   obtained   from  the calculation.    The  correct
description   of the measured angular   distributions  would require a
considerable dispersion   around   the mean   values, which   is   not
calculated by HICOL, and mean angles smaller by $\approx 30^{\rm o}$.

An  upper  limit to  the  reaction  time  can  be estimated using  the
characteristic rotational  time of the system  for the longest lasting
trajectories.   At  E$_{lab}=418$~MeV,  for  the average  quasi-fission
trajectory,   contributing in  the   symmetric mass   bin A=$120-140$,
L=$46\hbar$ and moment of inertia of two  touching spheres we estimate
$\tau \sim 30\times 10^{-21}$s. In HICOL, the rotational frequency and
the moment of  inertia are changing along  the  trajectory and reflect
the evolution of the shape and rotational degrees of freedom.  HICOL's
result for the same trajectory is $\tau=15\times 10^{-21}$~s, which is
in  excellent agreement with  the   time  extracted from the   neutron
measurement,  $\tau =  18\pm 9  \times   10^{-21}$~s \cite{Hin92}.  In
contrast  Wilczynska's  calculation \cite{Wil95},  extracted   $\tau =
80^{+55}_{-25}    \times  10^{-21}$~s     from     the  same     data.
Table~\ref{t:table3} gives the HICOL results  obtained for the average
trajectories  contributing  in  symmetric  quasi-fission for  the  two
energies measured. 
                              
The two  completely  different physics  approaches of Ref.\cite{Hin92}
and   this work  have  yielded similar  results for the   reaction
time scale, which indicates  that the one-body  friction implemented in
HICOL  gives   a     satisfactory   description    of    the  reaction
mechanism.  However,  a  dynamical  calculation  of  the  pre-scission
particle emission is necessary,  in  order to relate  the pre-scission
neutron multiplicity to the viscosity. Such  calculations were done in
this  work   and are  presented in  Appendix A. After adjusting the
statistical model parameters, so  that both the measured  pre-scission
neutron multiplicities     and mean  neutron    kinetic  energies  are
reproduced, the time scale derived from our dynamical calculation is in
agreement with the original ``static'' results of Ref.\cite{Hin92}. 

\section{One-body Dissipation Limits}
 
To determine the experimental limits to the one-body viscosity, 
implemented in HICOL, calculations were done using scaling factors to
the friction tensor ranging  from $k_{s}=0.5$ to $k_{s}=4$  and tested
versus the experimental mass and angular distributions.  These factors
were  kept  constant along  the  whole  trajectory, thus preserving
the original shape dependence of friction, which would be destroyed if
different scaling factors were used in different parts of the
trajectory.

The results   of the calculations   are  plotted in Fig.\ref{f:fig12}.
Scaling  factors   $k_{s}\geq~1.7$  lead to  scattering   angles $\geq
180^{o}$    and  orbiting   type  angular     distributions, which are
inconsistent with the data. In the lower limit, $k_{s}\leq~0.8$, there
is considerable discrepancy  between  the calculated and  the measured
mass distributions. This  analysis sets the  upper and  lower limits to
the one-body friction scaling factors for the temperature range of the
quasi-fission reactions studied in this work.  The two measurements in
the  present study  cover the  temperature range  of the  pre-scission
neutron data  from Ref.\cite{Hin92}.  In  contrast to the calculations
in  Ref.  \cite{Wil96} which found   that scaling factors $k_{s}=4-12$,
varying  with   temperature, were needed   to    describe the data  in
Ref.\cite{Hin92}, we consider both our  measurements are consistent
with one-body viscosity. Supporting this conclusion, Feldmeier
in~\cite{Fel85} analyzed  the $^{238}$U$+^{48}$Ca at
E$_{lab}=$5.9~MeV/u data     from  Shen
et.al.\cite{She87},  and  reproduced the  measured  mass distributions
with   his    code.   A  pre-scission  GDR   $\gamma$-ray measurement
and analysis~\cite{Nes95} in the  quasi-fission reaction of 
$^{58}$Ni$+^{165}$Ho at E$_{lab}=368$~MeV was also consistent with 
one-body  friction.  

There  have   been   a number  of    studies  in   fusion-fission  and
fusion-evaporation  reactions based on  the  statistical model,  which
show  the need for dissipation in  the  fission degree of freedom. The
systematic studies by the Stony Brook group~(See Ref.~\cite{Hof95} and
references therein)  on pre-scission GDR $\gamma-$~rays indicates that
viscosity increases  with temperature and  is larger than the one-body
value.  Some  of these   results  were used  in  Ref.~\cite{Wil96} and
together with  the   pre-scission  neutron  multiplicity  calculations
discussed above were interpreted  as  evidence for strong two-body  viscosity
setting in at temperature $T\leq 2$ MeV. 

On the  other    hand,  other  studies  from   pre-scission   particle
multiplicities        (e.g.Ref.~\cite{Les93,Les93a})     and        GDR
$\gamma$-rays(e.g.Ref.~\cite{vtH96}) in the  same temperature range do
not  find evidence for such  large viscosity.  A recent measurement of
Morton et.al.~\cite{Mor97} of GDR  $\gamma$-ray multiplicities for the
$^{32}$S+$^{nat}$W,$^{208}$Pb reactions also found a delay time consistent
with  one-body  dissipation  and  $\gamma$-fission angular correlations
that could  be described without temperature  dependent viscosity.  At
present, the  dissipation mechanisms  in fusion-fission  reactions and
the temperature at  which the two-body dissipation sets  in is still a
subject  of considerable  debate. 

\section{Conclusions}

The present study concentrated on the quasi-fission reaction dynamics,
exploiting the unique observables available in this type of reactions:
mass-angle  correlations and mass distributions.   It aimed to clarify
and   reconcile the data and the   theoretical descriptions of nuclear
viscosity that  were  available   prior to  this.  We  find   that two
different   experimental   observables:   mass-angle     correlations,
pre-scission neutron multiplicities  have yielded similar  results for
the reaction time scales.  Our data  analysis in terms of the one-body
dissipation  theory was  qualitatively successful.   Our own dynamical
evaporation calculation with the  code  DIFHEAT was performed for  the
neutron   data (See Appendix  A)   and  reconciled with the   original
statistical  model  analysis of Hinde  et.al..   In  this  work it was
shown, how the mass-distributions by themselves  are also sensitive to
the dissipation mechanism. The new results do not confirm the onset of
two-body viscosity in the temperature range below T=2.5 MeV. 
              
\acknowledgments
The authors would like to express their gratitude to the staff at the 
Nuclear Structure Laboratory in Stony Brook for their continuous
support. 
The targets were skillfully prepared by A. Lipski.
Fruitful discussions with P. Paul, B.B. Back and H.W. Wilschut
are gratefully acknowledged.

This work was sponsored in part by the U.S. National Science
Foundation.

I.Di\'{o}szegi acknowledges the support of OTKA Foundation, grant no 017285.
\appendix
\section{HICOL plus Evaporation}
\label{ss:evap}

In the Wilczynska  et.al.\cite{Wil95} approach, HICOL's trajectory was
divided into time  steps. A statistical  model calculation was applied
for each step, replacing the excitation energy with HICOL's dissipated
energy minus the energy lost in evaporation from the previous step. 

A similar  code, was  developed independently  in  this work   using a
somewhat different approach.   Instead of dividing the trajectory into
time steps, a  continuous evaporation  process was assumed,  following
Refs.\cite{Cha91,Del86}. The results of Refs.~\cite{Wil95,Wil96} could
be reproduced with our code DIFHEAT. 

Calculating the accumulated neutron multiplicity along the L=40$\hbar$
trajectory in $^{64}$Ni$+^{197}$Au at E$_{lab}=$418 MeV, using a level
density  parameter  $a=A/9$~MeV$^{-1}$ (the value used  in the
neutron measurement),  one  finds  that  the  calculated  pre-scission
multiplicity  does   not   reach    the  experimental   value.      In
Refs.~\cite{Wil95,Wil96}, the authors concluded that the reaction time
predicted by HICOL using one-body friction is too short  to be able to
accommodate the measured pre-scission neutron multiplicities. 

\begin{figure}
\begin{center}
\centerline{\psfig{file=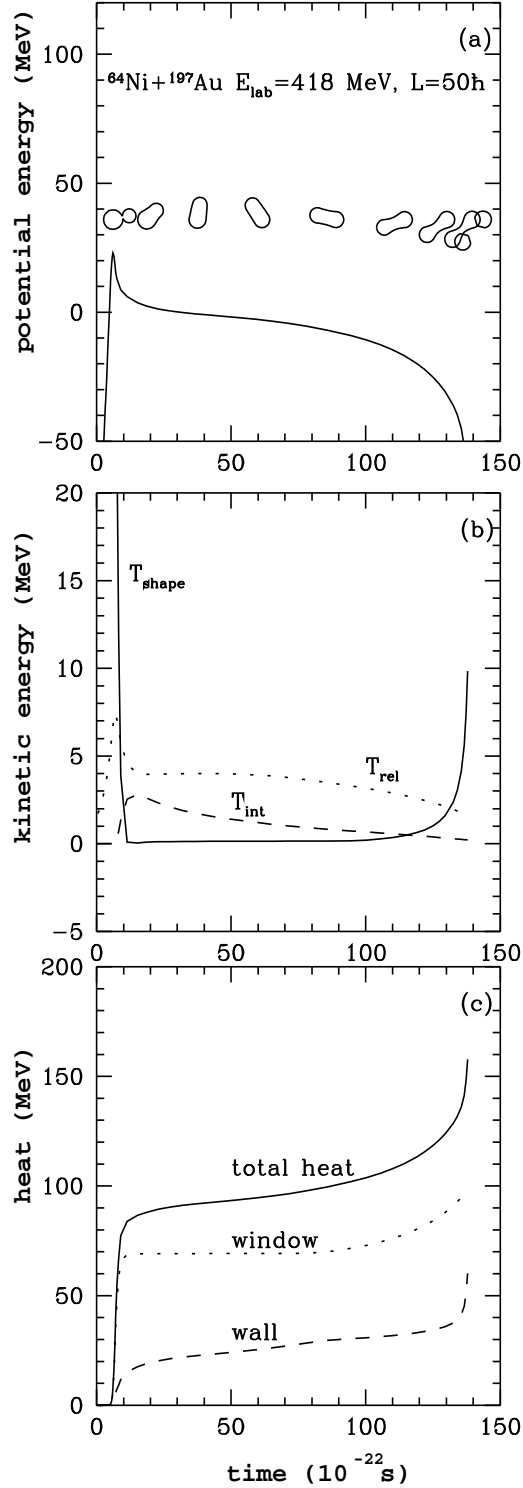,height=20cm,angle=0}}
\end{center}
\vskip -1 cm
\caption{Along the trajectory calculations:(a) potential energy and shapes.
(b)Kinetic energy in the shape degrees of freedom T$_{s}$, relative
rotation T$_{rel}$ and intrinsic rotation T$_{int}$. (c) 
Dissipated energy: total heat and the relative contributions from  
the window and wall dissipation.\label{f:fig6}}
\end{figure}

\begin{figure}
\begin{center}
\centerline{\psfig{file=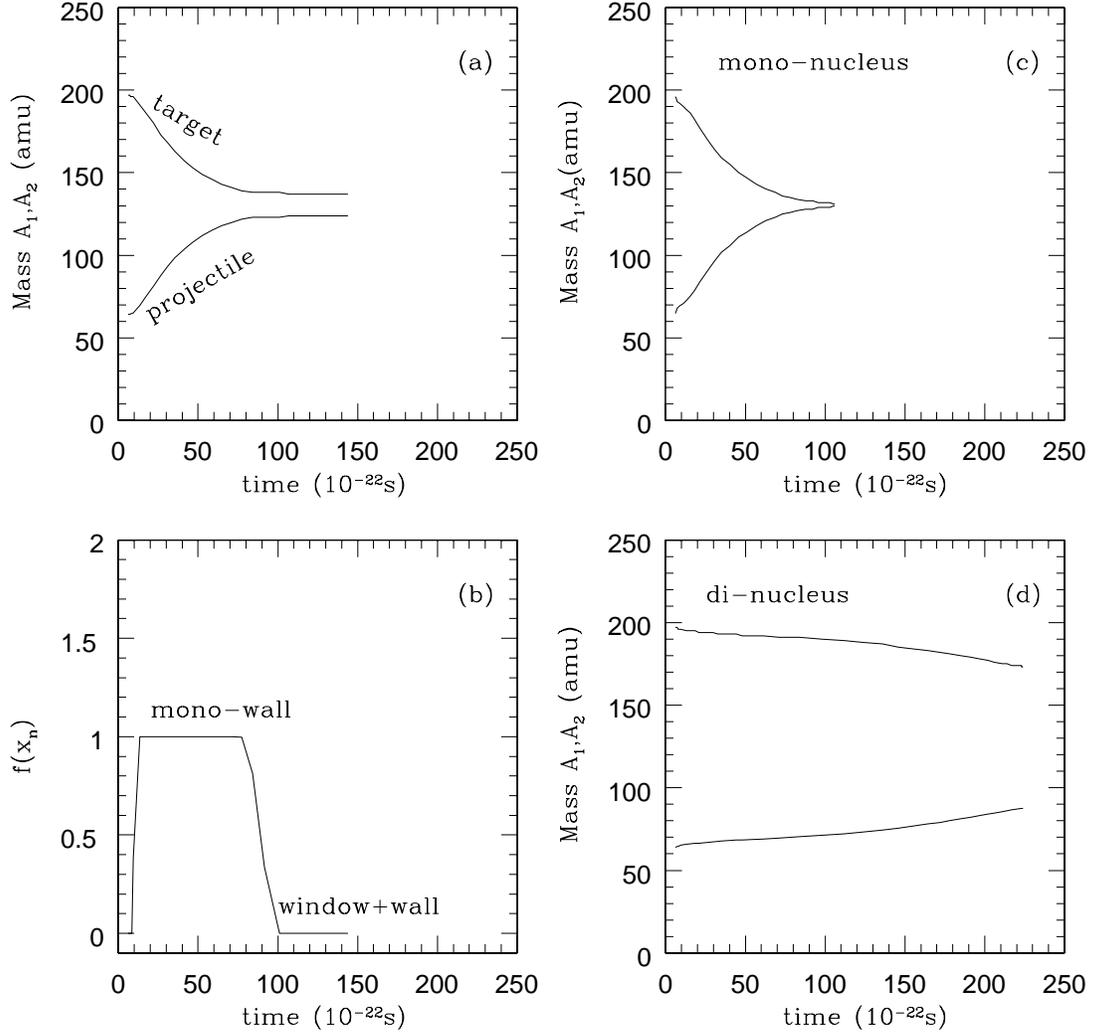,height=15cm}}
\end{center}
\vskip -1 cm
\caption{Window-plus-wall or mono-wall? (a) The mass evolution of the
projectile-like and the target-like fragments is calculated for the
E$_{lab}=418$ MeV, L=50$\hbar$ trajectory with window-plus-wall
and mono-wall friction mixed with a from factor $f(x_{n})$ as described
in the text. (b) The evolution of the form factor $f(x_{n})$ along the
same trajectory. (c) Fragment mass evolution calculated without the
window friction ( $f(x_{n})=1$ along the whole trajectory independent
of the shape) and (d) the mass evolution with $f(x_{n})=0$ -
window-plus-wall everywhere.\label{f:fig7}}
\end{figure}

\begin{figure}                  
\begin{center}
\centerline{\psfig{file=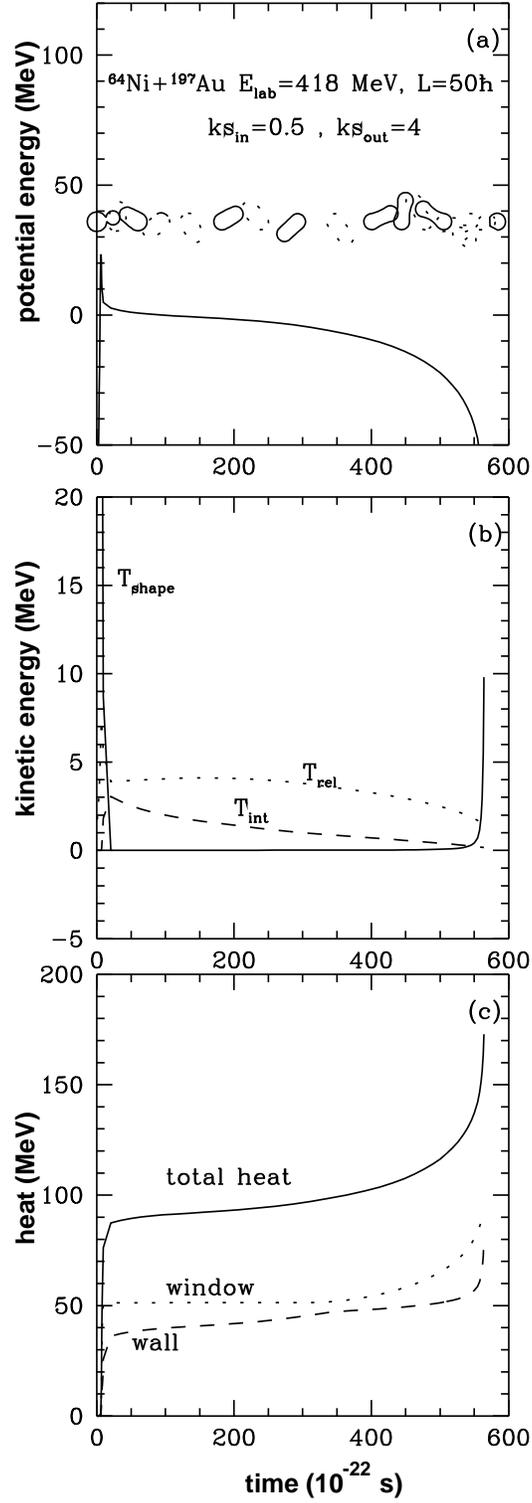,height=20cm,angle=0}}
\end{center}
\vskip -1 cm
\caption{As per Fig.\ref{f:fig6} but calculated with large
friction as in Ref.[11].\label{f:fig8} }
\end{figure}

\begin{figure} 
\begin{center}
\centerline{\psfig{file=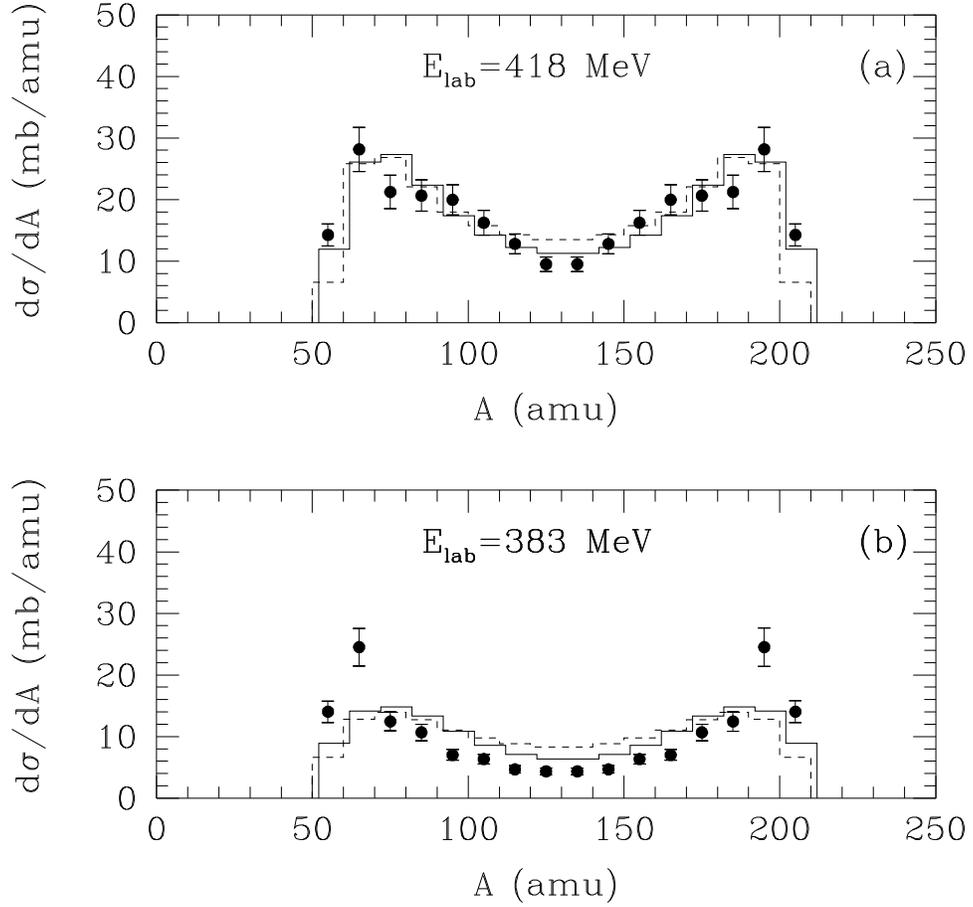,height=12cm,angle=0}} 
\end{center}
\caption {Mass distributions:(a) E$_{lab}=418$~MeV and 
(b) E$_{lab}=383$~MeV. The data are shown with
points. The histograms give HICOL calculations: dashed lines - unmodified
HICOL; solid lines - the window-plus-wall to mono-wall form factor is modified
as described in the text.\label{f:fig9} }
\end{figure}

\begin{figure}
\begin{center}
\vskip -2cm
\centerline{\psfig{file=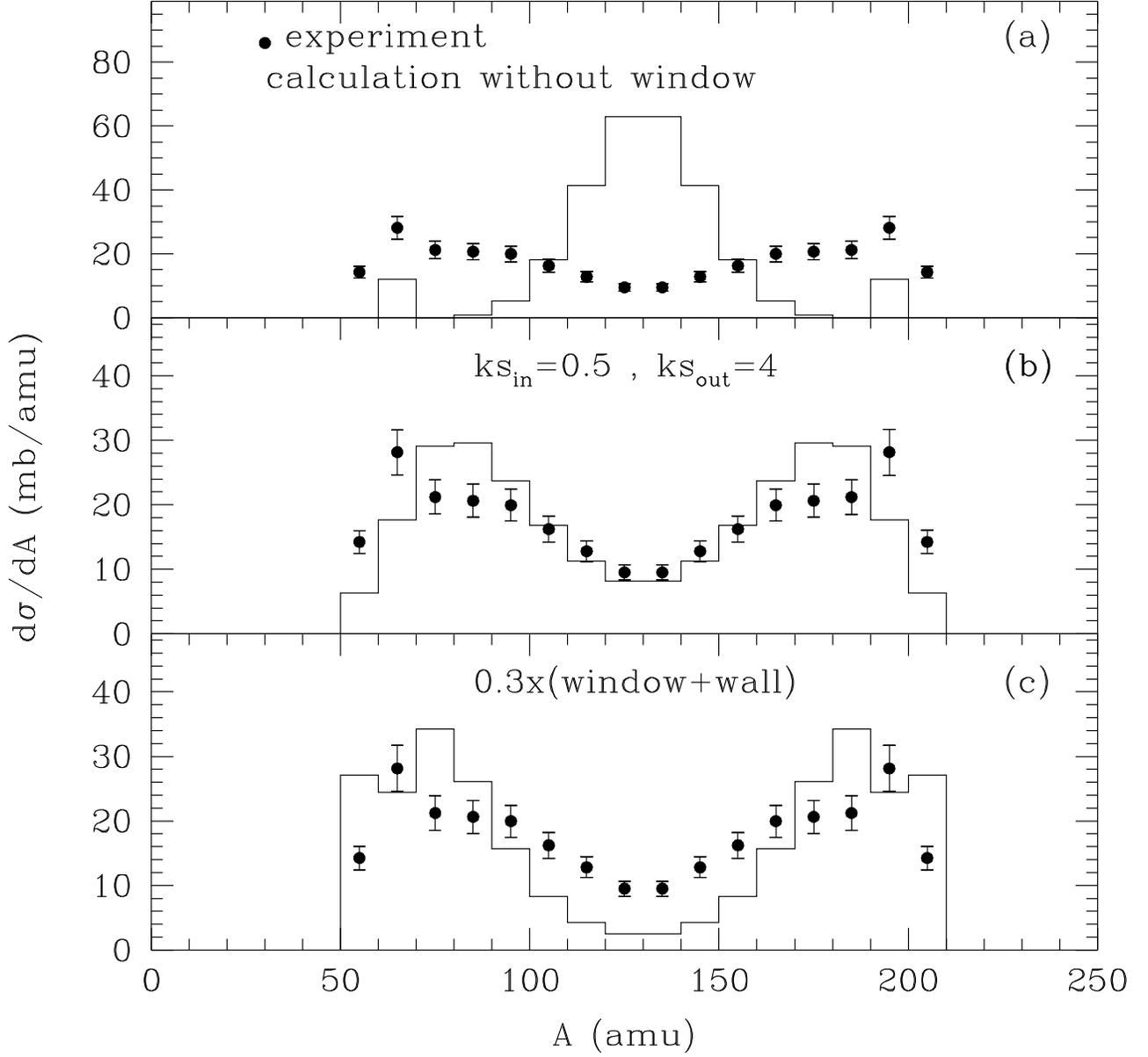,height=16cm}}
\end{center}
\caption {Mass distribution calculations for E$_{lab}=418$~MeV in
comparison to the data: (a) a calculation without the window friction;
(b) a calculation  with friction scaling according to Wilczynski; 
(c) $0.3\times$(window+wall) along the whole trajectory. \label{f:fig10}}
\end{figure}

\begin{figure} 
\begin{center}
\centerline{\psfig{file=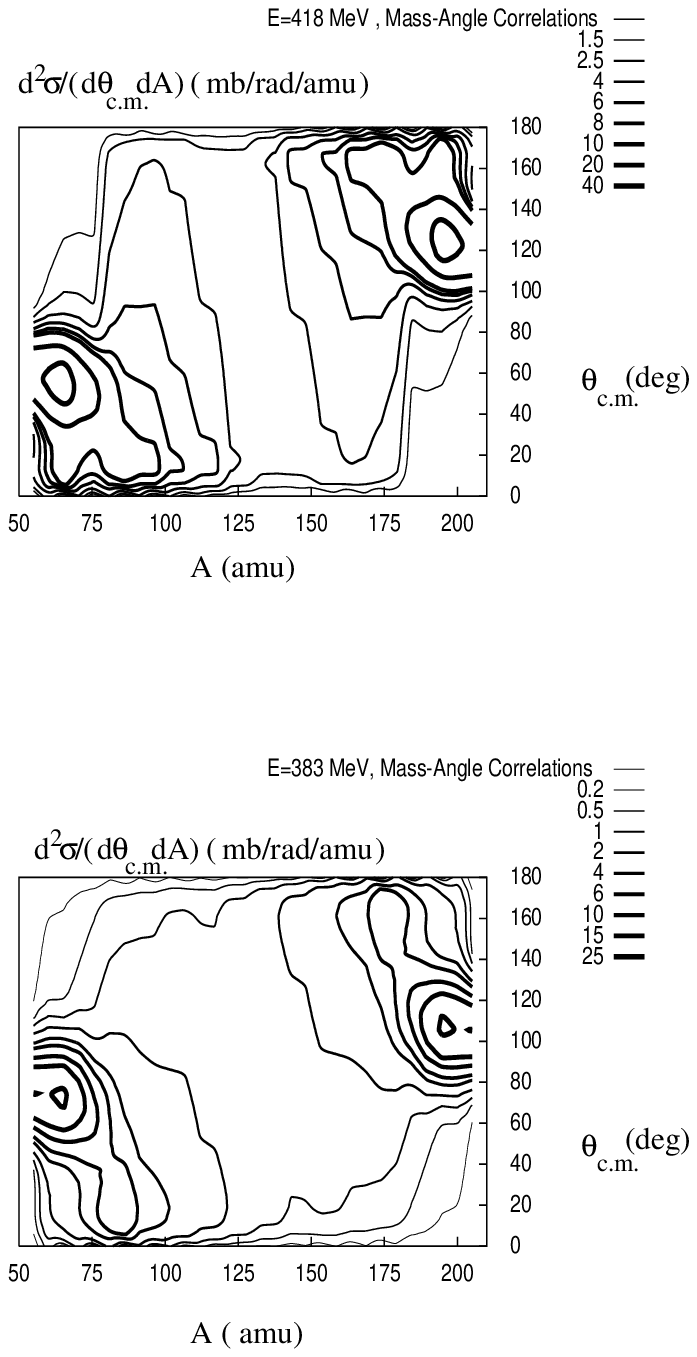,height=18cm,width=12cm,angle=0}} 
\end{center}
\caption{Double differential cross-sections
$d^{2}\sigma/(d\theta_{cm}dA)$
(mb/rad/amu) from the two energies measured. The contour levels are shown in
the plots.\label{f:fig11} }
\end{figure}

\begin{figure}
\begin{center}
\centerline{\psfig{file=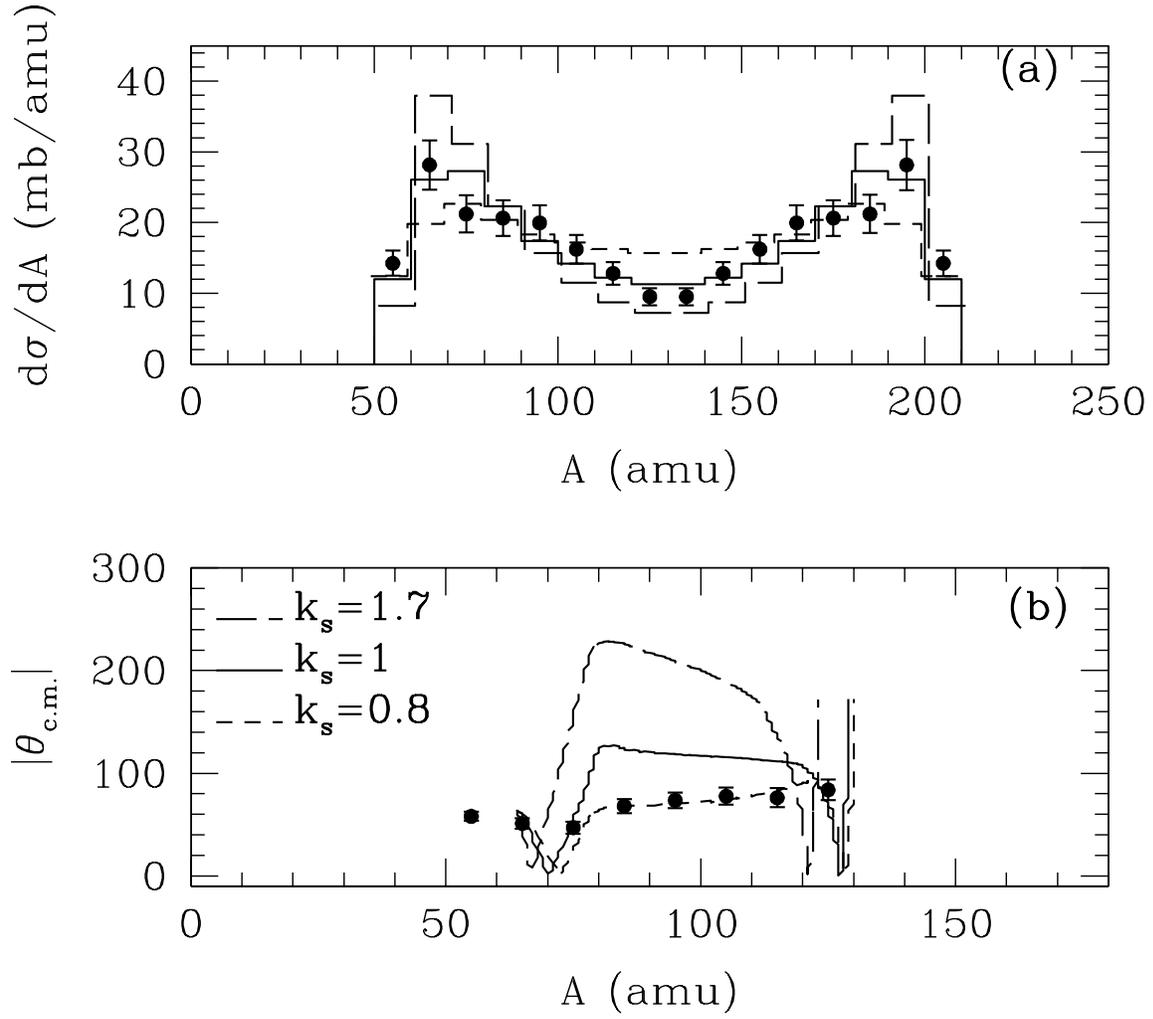,height=16cm,angle=0}}
\end{center}
\caption{Mass distributions (a) and mean scattering angles (b) calculated 
with HICOL with scaling factors : $ks=1.7$ - long-dashed histograms and curves,
$ks=0.8$ - short-dashed histograms and curves, $ks=1$ - full lines. The
experimental data for E$_{lab}=418$ MeV are shown with points.\label{f:fig12}}
\end{figure}

A different  approach to  reconcile the measured  pre-scission neutron
multiplicity with the dynamical trajectory  calculation was applied in
this   work.  It   appears    that the   discrepancy  between  Hinde's
et.al.\cite{Hin92}    and     Wilczynski's    et.al.\cite{Wil95,Wil96}
calculations does not   stem from the ``static'' versus  ``dynamical''
approach,  but   rather  reflects the   different  nuclear temperature
achieved in  the two cases.  In Hinde's  calculation the level density
parameter was   kept fixed at  the  value  $a=A/9$~MeV$^{-1}$, but the
excitation    energy    of the system      was    treated as   a  free
parameter\footnote{E$^{\ast}=$E$_{x}+\Delta$E$_{x}$.  E$_{x}=95.3$~MeV
was  determined from  the  $Q$  value  of  the reaction  and  $\Delta$
E$_{x}=50$ MeV was obtained  from the fit to  the data.}  and adjusted
in order to reproduce   {\it{both}} the measured  pre-scission neutron
multiplicity  and spectral shapes.   Although routinely in statistical
model calculations the level  density parameter  is varied, while  the
excitation energy is fixed, as Hinde pointed out, both observables are
sensitive to the   nuclear   temperature $T=\sqrt{E^{\ast}/a}$,  which
contains  the  ratio   of  the   two   quantities.    The authors   of
Ref.\cite{Wil95} argued that   the dynamically calculated   excitation
energy, as  predicted by HICOL \footnote{See  the  total heat curve in
Fig.\ref{f:fig6}(c).}  should  be   used.   Keeping  the  level density
parameter at the  same value  of  $a=A/9$~MeV$^{-1}$, but using  lower
excitation energy along the trajectory ($E^{\ast}\sim100$~MeV) results
in  lower nuclear  temperature.  Consequently, this calculation yields
longer neutron lifetimes and, in  addition, {\it{ does not}} reproduce
the mean neutron kinetic energy. 

\begin{figure} 
\vskip -2cm
\begin{center}
\centerline{\psfig{file=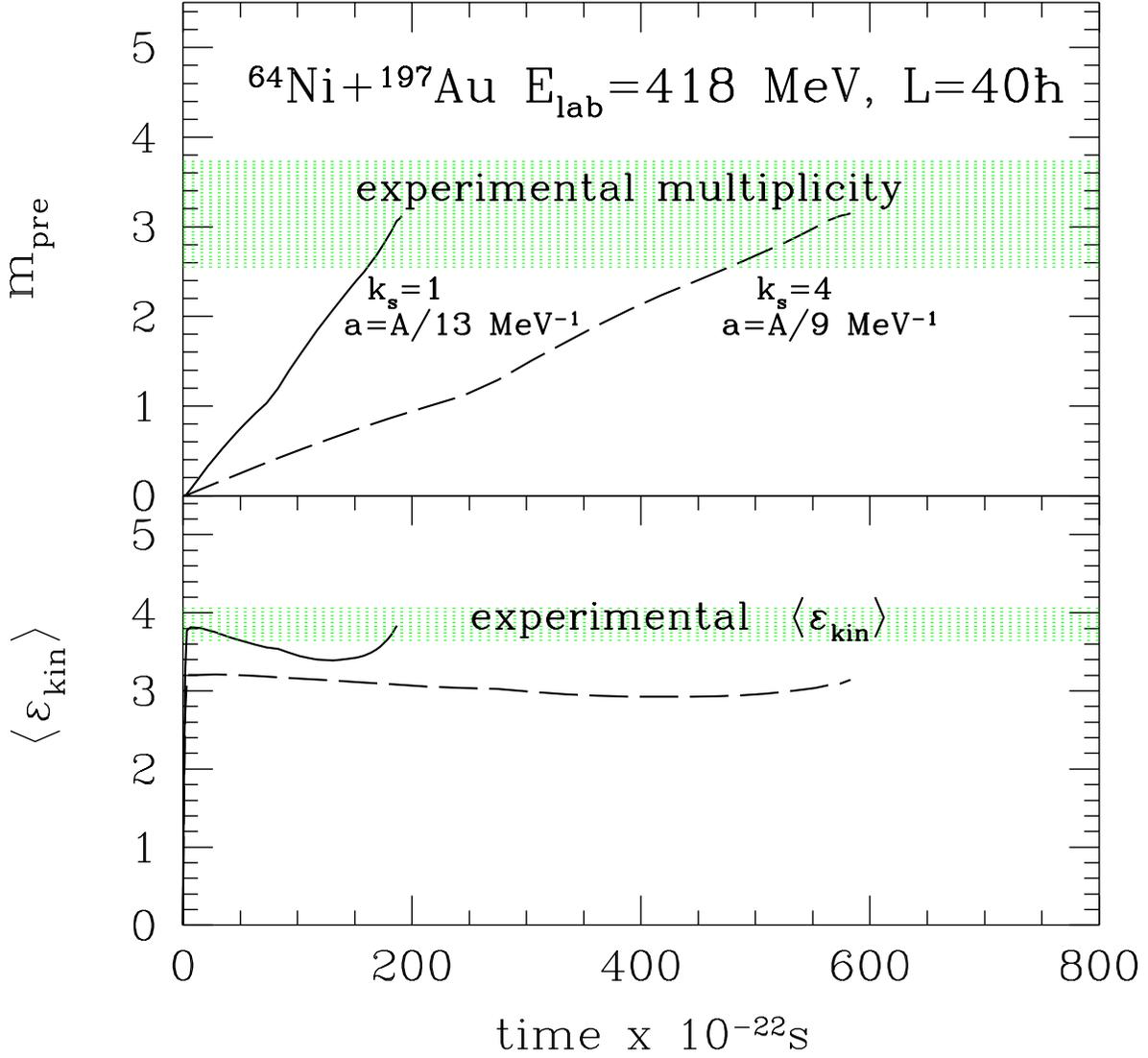,height=16cm}} 
\end{center}
\caption {Pre-scission neutron multiplicity and mean neutron
kinetic energy along the trajectory.
\label{f:fig13} }
\end{figure}

The dashed curves in Fig.~\ref{f:fig13}  show the neutron multiplicity
and the mean  neutron kinetic energy  along the trajectory, calculated
with $a=A/9$~MeV$^{-1}$  and an input heat  curve obtained  by scaling
the friction  in HICOL  by a  factor  $k_{s}=4$.  The   shaded regions
indicate the experimental  multiplicity and mean kinetic energy values
reported in Ref.~\cite{Hin92}. 

Since   HICOL indeed  provides  a more   consistent description of the
excitation    energy,  than     the   fitting  procedure      used  in
Ref.~\cite{Hin92},  in  order   to   produce    the  same  temperature
(T$\approx2.2$~MeV), as derived  from the neutron spectra measured  by
Hinde,   we   need   to vary   the    level   density parameter.  From
Fig.~\ref{f:fig6}(c) we  find  that  $E^{\ast}\approx100$ MeV, and  we
estimate that  $a=A/12.7$~MeV$^{-1}$ should reproduce  {\it both}  the
neutron lifetimes  and mean kinetic energy.    The solid lines  in the
Fig.~\ref{f:fig13} show the result from DIFHEAT, using a level density
parameter $a=A/13$~MeV$^{-1}$  and  a one-body dissipation  heat curve
coming from HICOL. 

Unlike the conclusion in   Ref. \cite{Wil96}, that large friction   is
necessary   to   reproduce  the   experimental   pre-scission  neutron
multiplicities, we  find  that the  reaction time  extracted from  the
neutron data~$\sim  20\times 10^{-21}s$   is consistent  with  HICOL's
predictions and with the evaporation calculation using DIFHEAT, if the
statistical model parameters  are properly constrained. While HICOL is
free of  adjustable parameters,  in any statistical  model calculation
one  needs  at least two  observables, to   be  able to  constrain the
multi-parameter   model calculation.  It  is beyond  the scope of this
work to investigate the influence  of all statistical model parameters
on   the  results from DIFHEAT.   There  is a  significant deformation
dependence both in the   particle transmission coefficients and  their
binding  energies \cite{Les91,Les93}.    Accounting for these  effects
might change the   value of the  level   density parameter needed   to
describe the  data. However, we  note that values  as small  as $a\sim
A/13$~MeV$^{-1}$ have been  previously deduced from particle  spectral
shapes in heavy hot systems \cite{Fin94,Fin94t}. 

\bibliographystyle{unsrt}
\bibliography{bib}

\newpage


\begin{table}
\begin{tabular}{ccccc} 
Reaction    & E$_{lab}$ MeV  &  $\nu_{pre}$   & $\tau_{f}\times
10^{-21}s$ \tablenotemark[1] & $\tau_{f}\times10^{-21}s$ \tablenotemark[2]  \\
\tableline
\\
$^{40}$Ar$+^{238}$U   &  249 & 3.25$\pm0.2$ & 30$\pm10$ & 390$\pm_{60}^{80}$\\
\\
$^{64}$Ni$+^{197}$Au  &  418 & 3.15$\pm0.6$ & 18$\pm9$  & 80$\pm_{25}^{55}$ \\ 
\\
$^{64}$Ni$+^{208}$Pb  &  418 & 3.25$\pm0.6$ & 28$\pm10$ & 70$\pm_{25}^{35}$ \\ 
\\
$^{64}$Ni$+^{238}$U   &  418 & 4.00$\pm0.8$ & 15$\pm8$  & 120$\pm_{40}^{90}$ \\

\end{tabular}
\tablenotetext[1]{from Ref.\ \cite{Hin92}}
\tablenotetext[2]{from Ref.\ \cite{Wil95}}
\caption{Quasi-fission reactions, pre-scission neutron multiplicities and
reaction times.\label{t:table1}}
\end{table}

\newpage
\begin{table}
\widetext
\begin{tabular}{c|ccc|ccc} 
&\multicolumn{3}{c|}{ E$_{lab}=418$ MeV}&
\multicolumn{3}{c}{ E$_{lab}=383$ MeV} \\
\tableline
\tableline
Mass  & $\sigma_{c}$ & $\sigma_{DI}$ & $\sigma_{touch}$ 
& $\sigma_{c}$ & $\sigma_{DI}$ & $\sigma_{touch}$ \\
(amu) & (mb) & (mb) & (mb) & (mb) & (mb) & (mb) \\
\tableline
 
120-130  & $95\pm 21$ & & $95\pm 21$ & $43\pm 10$ & & $43\pm 10$ \\ 

110-120 & $128\pm 29$ & & $128\pm 29$ & $46\pm 11$ & & $46\pm 11$ \\

100-110 & $162\pm 37$ & & $162\pm 37$ & $63\pm 15$ & & $63\pm 15$ \\

90-100 & $199\pm 45$ & & $199\pm 45$ & $70\pm 16$ & & $70\pm 16$ \\

80-90 & $206\pm 47$ & & $206\pm 47$ & $106\pm 24$ & & $106\pm 24$ \\

70-80 & $137_{-68}^{+34}$ & $75_{-19}^{+38}$ & $212\pm 59$
& $62_{-31}^{+15}$ & $62_{-15}^{+31}$ & $124\pm 32$ \\       

60-70 & $95_{-48}^{+24}$ & $186_{-46}^{+92}$ & $281\pm 79$
& $50_{-25}^{+12}$ & $194_{-49}^{+97}$ & $245\pm 65$ \\      

50-60 &$29_{-15}^{+7}$&$113_{-28}^{+56}$&$142 \pm 40$
& $6.3_{-3.1}^{+1.5}$ & $134_{-33}^{+67}$ & $140\pm 36$ \\   
\tableline

total & $1053_{-350}^{+263}$ & $374_{-93}^{+190}$ & $1427\pm 357$ 
& $450_{-194}^{+135}$ & $391_{-117}^{+194}$ & $841\pm 210$ \\
\end{tabular}
\caption{Angle-integrated cross-sections in bins of 10 mass units from the two
energies studied.\label{t:table2}}
\end{table}

\newpage
\begin{table}
\narrowtext
\begin{tabular}{ccc} 
E$_{lab}$~(MeV)	&  $\langle L \rangle$~$\hbar$	& time$\times 10^{-21}$~s\\
\tableline
418	&	46	&	15\\
383	&	22	&	16\\
\end{tabular}
\caption{Reaction times for the average quasi-fission trajectories
contributing in the A$=120-140$ mass range.\label{t:table3}}
\end{table}

\newpage
\widetext
\newpage

\end{document}